\documentclass{article}
\usepackage{graphicx}
\title{Accurate energy spectrum for double-well potential: periodic basis}
\author{P. Pedram$^a$\thanks{Email: pouria.pedram@gmail.com}, M. Mirzaei$^b$ and S. S. Gousheh$^b$
\\ {\small $^a$Plasma Physics Research Center,
Science and Research Branch, Islamic Azad University, Tehran,
Iran}\\{\small $^b$Department of Physics, Shahid Beheshti
University, Evin, Tehran 19839, Iran}}
\begin{document}
\maketitle \baselineskip 24pt
\begin{abstract}
We present a variational study of employing the trigonometric basis
functions satisfying periodic boundary condition for the accurate
calculation of eigenvalues and eigenfunctions of quartic double-well
oscillators. Contrary to usual Dirichlet boundary condition,
imposing periodic boundary condition on the basis functions
results in the existence of an inflection point with vanishing
curvature in the graph of the energy versus the domain of the
variable. We show that this boundary condition results in a higher
accuracy in comparison to Dirichlet boundary condition. This is due
to the fact that the periodic basis functions are not necessarily
forced to vanish at the boundaries and can properly fit themselves
to the exact solutions.
\end{abstract}

\textit{Keywords}: Linear variations; Double-well potential;
Eigenvalues; Eigenfunctions; Basis sets.

\section{Introduction}\label{sec1}
\begin{figure}
 \centering
\includegraphics[width=10cm]{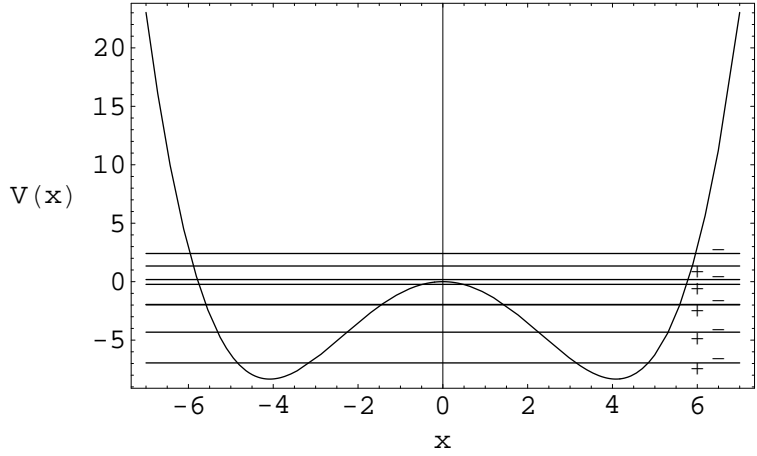}
\caption{One-dimensional double-well potential where $k=1$ and
$\lambda=0.03$. The $\pm$ signs on each energy level refers to the
parity of the corresponding eigenstate. Starting from the ground
state up, the energy splitting between the levels below the cusp are
approximately \{$4.3\times 10^{-6}$, $6.1\times 10^{-4}$, $3.0\times
10^{-2}$\}.}\label{Fig1}
\end{figure}

The phenomena of quantum tunneling in the double-well potential
(Fig.~\ref{Fig1})
\begin{equation}\label{main}
V(x)=-kx^2+\lambda x^4\hspace{3cm}(k,\lambda>0)
\end{equation}
and finding the energy eigenvalues and its stationary states is a
long-standing and well-known problem in quantum mechanics. The
interest in this problem ranges from various branches of physics to
chemistry. After the seminal works by Bender and Wu \cite{bender},
numerous papers have been appeared in the literature to tackle this
problem. In spite of its simple form, this problem presents a
non-trivial model in both quantum mechanics and quantum field
theory. In fact, this model can be considered as $0+1$-dimensional
quantum field theory $\lambda \phi^4$ which has no spatial
dimensions.

In chemistry, this type of potential corresponds to two equilibrium
positions or molecular configurations. The periodic inversion of
ammonia ($NH_3$) is a well-known example of quantum tunneling. The
internal rotation in $CH_3CH_3$ molecule from one configuration to
another is a good example of a symmetric double-well potential. On
the other hand, the asymmetric $H_2O_2$ hindered rotor or beryllium
dicyclopentadienyl represent asymmetric double well potentials.
Ammonia ($NH_3$), cyanamide ($NH_2CN$), $PH_3$, and $AsH_3$ are
chemical examples of symmetrical double well potentials where atoms
can tunnel through the barrier.

It is obvious that it is not possible to set up a manageable
perturbation theory even for an infinitesimal value of the coupling
strength $\lambda$ to calculate the exponentially small energy
splitting between quasi-degenerate levels in a double-well. This
stems from its asymptotic divergent nature \cite{landau} due to the
eventual dominance of the perturbative correction over the
unperturbed contribution for large amplitude of oscillation
\cite{bender}. This has been the driving force behind the
development of non-perturbative methods for this kind of problems.
Among them are the semiclassical approximation \cite{slavyanov},
finite-difference technique \cite{Witwit}, hypervirial
recurrencerelation scheme \cite{Hodgson},  renormalized
hypervirial-Pad\`{e} scheme \cite{Bansal}, variational matrix
solution \cite{Balsa}, Rayleigh-Ritz variational method supplemented
by the Lanczos algorithm \cite{Bishop}, Brillouin-Wigner
perturbation theory based on shifted oscillator variational
functions \cite{Arias}, instanton method \cite{Kleinert,Gildener},
transfer matrix method \cite{Zhou}, uniform approximation of the
logarithmic derivative of the ground state eigenfunction
\cite{turbiner}, and many other specific methods
\cite{Banerjee,Bhattacharyya3,Fack,methods}. For the double-well potential, even
quite sophisticated techniques, such as the renormalized
hypervirial-Pad\`{e} scheme \cite{Bansal}, fail to calculate the
semi-degenerate low energy states.

The WKB approximation is widely used for its simple mathematical
form, but the results are known to be inaccurate due to its inherent
defect in the connection formula. One can take the quadratic
connection formula instead of those related to the Airy functions to
modify the WKB result for the ground state \cite{Gildener,Carlitz}.
Some other refinements have been developed to improve the accuracy
of WKB by changing the phase loss at the classical turning points
\cite{Friedrich,Park}. The anharmonicity is also taken into
consideration in the case of small separation distance between the
wells \cite{Park}.

In the instanton method, one uses imaginary time path integral to
obtain the classical action the so-called Euclidean action.
Qualitatively, this method is useful for understanding the quantum
tunneling, which has no classical counterpart. However, the
quantitative calculation of the energy splitting in the double-well
potential by this method is inaccurate because the Euclidean
propagator can be obtained only in the limit of infinite separation
between the two potential minima \cite{Kleinert}, which corresponds
to zero tunneling probability. Thus, the validity of the instanton
approach is restricted to the case in which the two potential minima
are far apart, and its accuracy is expected to be reduced as they
become close.

Various variational methods are usually used in quantum mechanics,
statistical mechanics and field theory. In quantum mechanics,
variational parameters are incorporated into trial wave functions
and trial Hamiltonians. The Rayleigh-Ritz method is the minimization
of the ground state energy with respect to these variational
parameters. The applications of this formalism using various set of
basis such as harmonic-oscillator \cite{14}, Chebyshev polynomials
\cite{17}, hypervirial theorems \cite{18}, and the coherent-states
\cite{19}, have been already appeared in the literature.

In Refs.~\cite{taseli,taseli2,Bhattacharyya,Bhattacharyya2}, it was
shown by numerical results that the eigenfunctions which obey
Dirichlet boundary condition
\begin{equation}\label{Dirichlet}
\Psi(-L) = \Psi(L) = 0,
\end{equation}
can be effectively used to find the spectrum of an unbounded
problem. Then, a critical distance $\hat L$ was defined, and it was
shown that the low lying energy levels $E_n(L)$ are equal to those
of $L=\infty$ with high accuracy, if the boundedness parameter $L$ is
in near vicinity of $\hat L$. In these works, they have employed
the Rayleigh-Ritz variational method and have chosen the
trigonometric functions (particle-in-a-box) as basis set, in order
to solve double-well potential and some other anharmonic symmetric
oscillators \cite{taseli,taseli2,Bhattacharyya,Bhattacharyya2}. They
have shown that the trigonometric functions are a suitable basis-set
in variational calculations in order to obtain highly accurate
results. Moreover, the model is simple, fast-convergent, efficient
and works for different kind of potentials
\cite{taseli,taseli2,Bhattacharyya,Bhattacharyya2}.

In this paper, we will also use trigonometric basis functions, but
with periodic boundary condition
\begin{equation}\label{Periodic}
\Psi(-L) = \Psi(L).
\end{equation}
In this case, contrary to Dirichlet boundary condition
(\ref{Dirichlet}), $E_n(L)$ has not any \emph{relevant} minimum value. In
fact, for periodic boundary condition $E_n(0)$ is zero, whereas
$E_n(0)$ is infinite for Dirichlet boundary condition. Note that,
imposing periodic boundary condition on  \emph{e.g.} the ground state
wave function, makes it more accurate than imposing Dirichlet
boundary condition. This is due to the fact that the Dirichlet
boundary condition enforces the wave function to vanish at $x=\pm L$,
but the periodic boundary condition lets the wave function to fit
itself more closer to the exact solution which is not necessarily
zero at the boundaries. Therefore, we show that using periodic
boundary condition, the same accuracy can be obtained with the
smaller number of basis.

The paper is organized as follows: In Sec.~\ref{sec2}, we outline
the general calculation scheme using the periodic boundary
condition. In Sec.~\ref{sec3}, we apply this method to the case of
double-well potential and compare our results with the previous
near-exact ones. For the case $k=1$, we find the accurate energy
eigenvalues for various value of $\lambda$ and obtain the optimum
value of $L$ in each case. Also, we outline the scaling properties
of the Hamiltonian to generalize the results to more complicated
forms of the potential. In Sec.~\ref{sec4}, we state our
conclusions.

\section{The variational method}\label{sec2}
Let us consider the following dimensionless time-independent
one-dimensional Schr\"{o}dinger equation
\begin{equation}\label{Schrodinger}
\left[-\frac{d^2}{dx^2}+V(x)\right]\Psi(x)=E\Psi(x),
\end{equation}
where $V(x)=V(-x)$. Since the potential is an even function of $x$,
to avoid large matrices, we can use even and odd basis functions
separately
\begin{equation}
\phi_m^0(x)=\frac{1+\delta_{m0}(2^{-1/2}-1)}{\sqrt{L}}\cos\left(\frac{m
\pi x}{L}\right),\hspace{2cm}m=0, 1, ...,
\end{equation}
and
\begin{equation}
\phi_m^1(x)=\frac{1}{\sqrt{L}}\sin\left(\frac{m \pi
x}{L}\right),\hspace{2cm}m=1, 2, ....
\end{equation}
Now, we can expand the solution using these orthonormal basis sets
\begin{eqnarray}
\Psi^i(x)= \sum_{m} A_{m} \phi_m^i(x), \label{eqpsitrigonometric}
\end{eqnarray}
where $i=0,1$ denotes even and odd solutions, respectively. Also, we have the following expansion
\begin{eqnarray}
V(x) \Psi^i(x)= \sum_{m}B_m\phi_m^i(x),\label{eqV}
\end{eqnarray}
where $B_{m}$ are coefficients that can be determined once $V(x)$ is
specified. By substituting Eqs.~ (\ref{eqpsitrigonometric}) and
(\ref{eqV}) into Eq.\ (\ref{Schrodinger}) and using the linear
independence of the basis functions, we have
\begin{eqnarray}
\left(\frac{m \pi}{L}\right)^2 A_{m}+B_{m}=E\, A_{m}.\label{eqAB2}
\end{eqnarray}
Moreover, we can find $B_{m}$'s using Eq.~(\ref{eqV}) and
(\ref{eqpsitrigonometric}), namely
\begin{eqnarray}
B_{m,i}= \sum_{m',} C_{m,m'} A_{m'},
\end{eqnarray}
where
\begin{eqnarray}
C_{m,m'}=\int_{-L}^{L} \phi_m(x)\,\,\, V(x)\,\,\,\phi_{m'}(x)\,\, dx.
\end{eqnarray}
Therefore we can rewrite Eq.~(\ref{eqAB2}) as
\begin{eqnarray}
\left(\frac{m \pi}{L}\right)^2 A_{m}+ \sum_{m'} C_{m,m'}\,\,
A_{m'}=E\, A_{m}.\label{eqAC}
\end{eqnarray}
By selecting a finite subset of the basis functions, {\it i.e.}
choosing the first $N$, Eq.~(\ref{eqAC}) can be written as
\begin{eqnarray}
D\, A=E\, A. \label{eqmatrix}
\end{eqnarray}
Therefore, the eigenvalues and eigenfunctions of the Schr\"{o}dinger
equation (\ref{Schrodinger}) are approximately equal to the
corresponding quantities of the matrix $D$.

\section{Results}\label{sec3}
\begin{figure}
\centerline{\begin{tabular}{ccc}
\includegraphics[width=7cm]{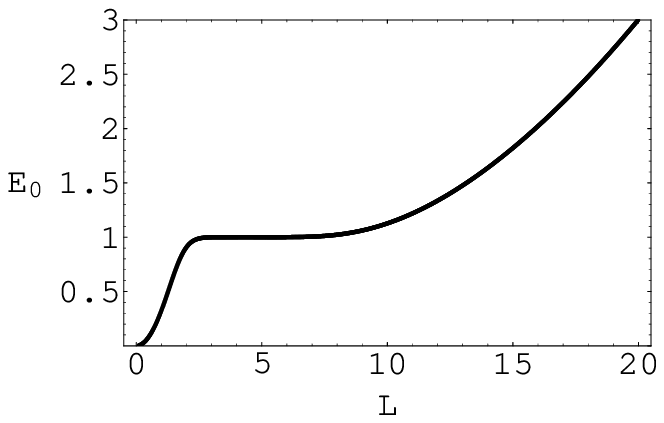}
\includegraphics[width=7cm]{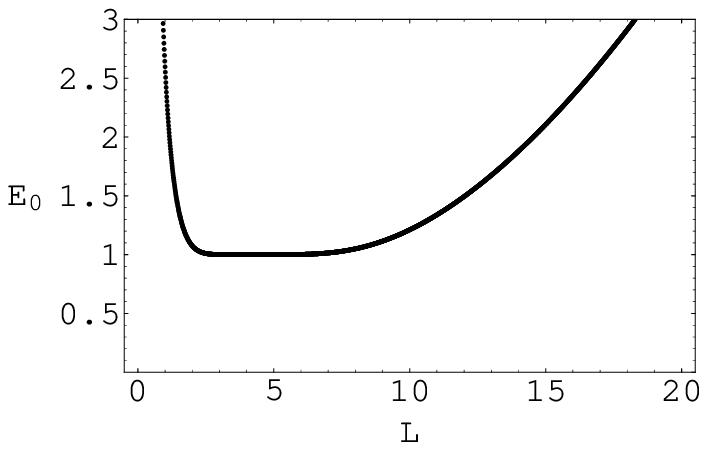}
\end{tabular}}
\caption{Ground state energy of Simple Harmonic Oscillator versus $L$ for $N=5$ using
periodic boundary condition (left) and using Dirichlet boundary
condition (right).}\label{figSHO}
\end{figure}

To establish the method, first let us study an exactly solvable case
where $k=-1$ and $\lambda=0$. Simple Harmonic Oscillator has a
well-known energy spectrum $E_n=2n+1$. Figure \ref{figSHO} shows its
ground state energy versus $L$ for both periodic and Dirichlet
boundary conditions. As apparent from the figure, for the case of
periodic boundary condition, the ground state energy starts from
zero at $L=0$ and grows as $L$ increases. Moreover, the graph shows
the existence of an inflection point around $L=5$ for $N=5$. In this
region, the evaluated eigenvalues are close to the exact one ($E_0=1$). Note
that, the position of the inflection point increases for a larger
set of basis functions. On the other hand, for Dirichlet
boundary condition, the graph of the ground state energy contains a
minima around $L=5$ for $N=5$. In this case, the optimal value of
$L$ ($\hat L$) is the position of this minima on $L$ axis. Also,
similar to the previous case, $\hat L$ increases for larger values
of $N$. This shows that the Rayleigh-Ritz variational formalism is only
applicable to the case of Dirichlet boundary condition.

\begin{figure}
\centerline{\begin{tabular}{ccc}
\includegraphics[width=7cm]{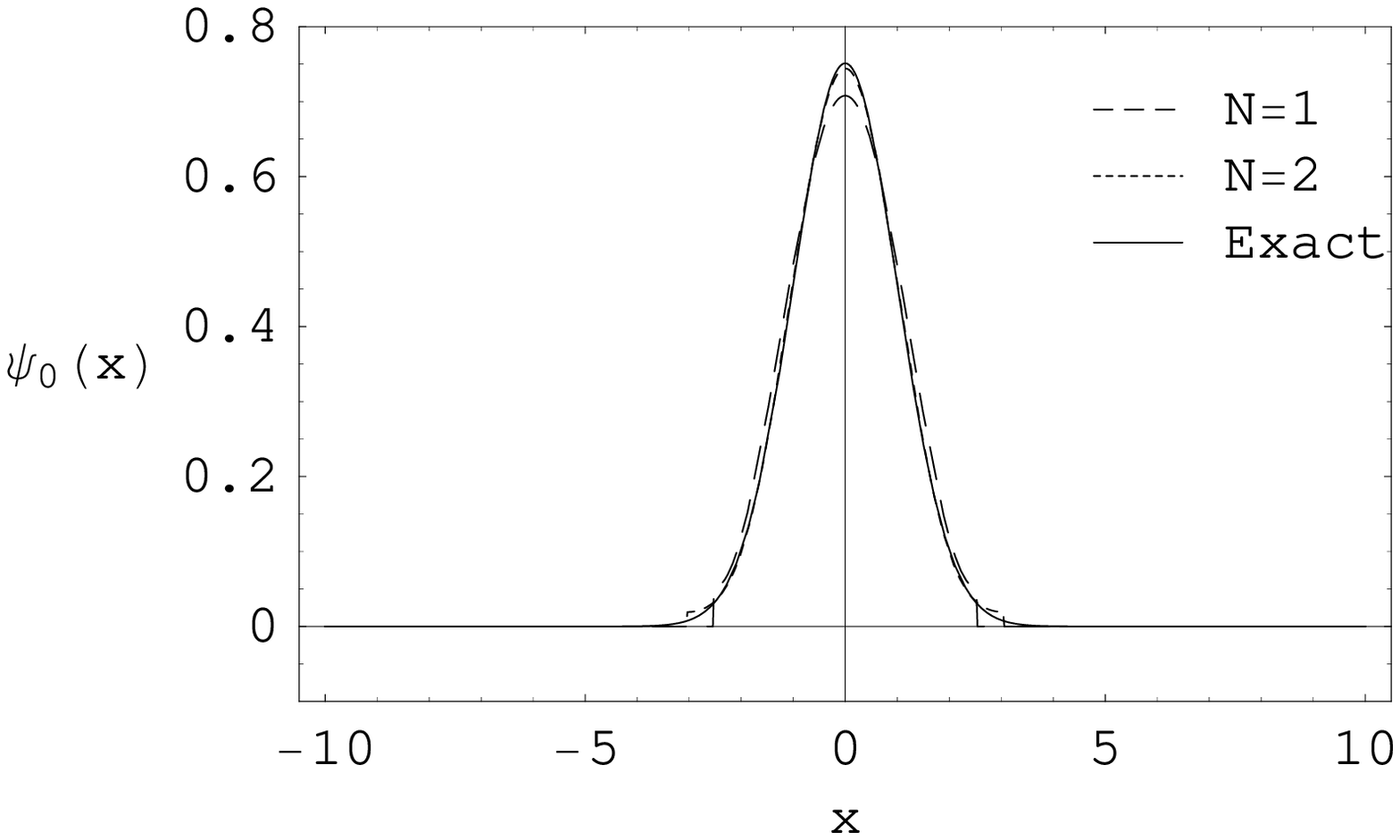}
 &\hspace{2.cm}&
\includegraphics[width=7cm]{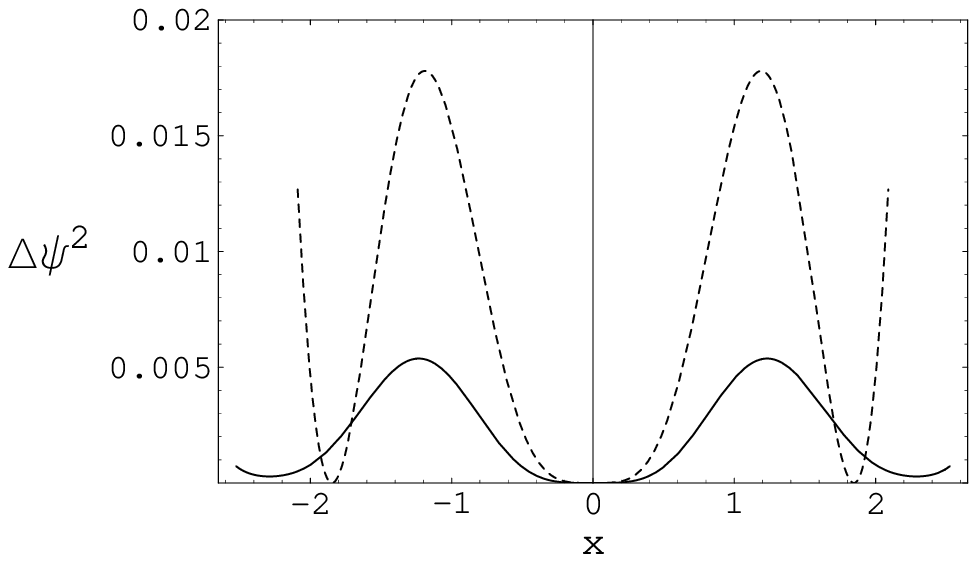}
\end{tabular}}
 \caption{Left, the exact and
approximate ground state wave functions of SHO for $\hat
L(1)=2.52479$ and $\hat L(2)=3.04635$. Note that for $N\ge2$ the
exact and approximate wave functions are practically
indistinguishable on the graph. Right, $\Delta
\Psi^2_{N=1}=|\Psi_{exact}-\Psi_{op}|^2$ versus $x$ for periodic
boundary condition (solid line) and Dirichlet boundary condition
(dashed line).}\label{figDelta}
\end{figure}

Figure \ref{figDelta} shows the exact and optimized wave functions
for $N=1,2$. In fact, for $N\ge2$ the exact and optimized wave
functions are practically indistinguishable on the graph. The
optimized ground state energy for $N=1$ using periodic boundary
condition is $1.008$ ($E_0^{exact}=1$) whereas this value for
Dirichlet boundary condition are $1.136$ and $1.006$ for $N=1$ and
$2$, respectively. Therefore, for this case, we need to utilize two
oscillatory terms satisfying Dirichlet boundary condition for
getting the same accuracy of one oscillatory term which satisfies
periodic boundary condition. Moreover, the optimized wave function
is also more accurate (Fig.~\ref{figDelta}). This discrepancy is
more apparent for larger values of $N$ which will be addressed for
the case of a double-well potential. As stated before, this is due
to the fact that for the periodic boundary condition, the ground
state wave function is not forced to vanish at $x=\pm L$ and can
better fit itself to the exact solution which certainly is not zero
there.

\begin{figure}
\centerline{\begin{tabular}{ccc}
\includegraphics[width=7cm]{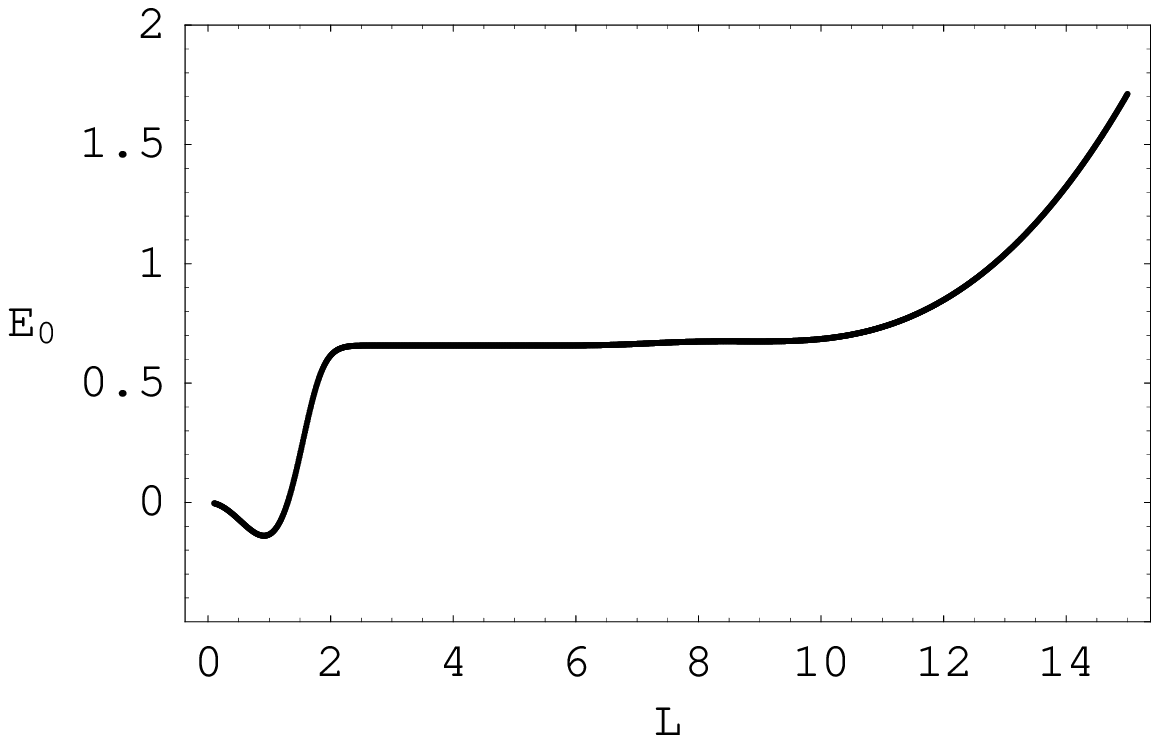}
 &\hspace{2.cm}&
\includegraphics[width=7cm]{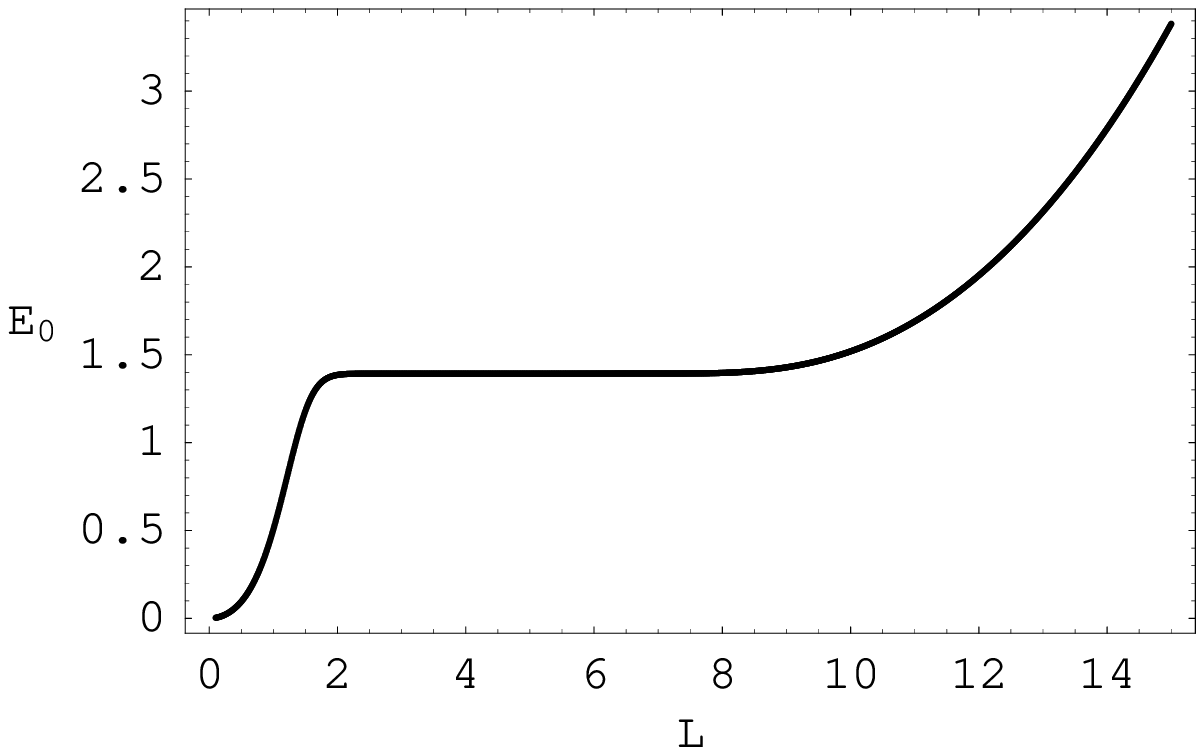}
\end{tabular}}
 \caption{Ground state energy versus $L$ for $N=8$, $\lambda=1$, $k=1$ (left), and $k=-1$ (right).}\label{NL}
\end{figure}

Now, let us consider the situation that the potential contains an
anharmonicity term ($\lambda>0$). In this case, the sign of $k$ can
be either positive (anharmonic oscillator) or negative
(double-well potential). In our variational formalism, contrary to Dirichlet
boundary condition (\ref{Dirichlet}), $E_n(L)$ has no minimum value
for $L>L_c$ (Fig.~\ref{NL}), where $L_c$ is the position of
classical turning point. Note that, since $\hat L$ should be much larger
than $L_c$ \cite{Bhattacharyya,Bhattacharyya2}, we consider the
behavior of $E_n(L)$ only for $L>L_c$. Figure \ref{NL} shows the
behavior of the ground state energy for a double-well potential
$V(x)=-x^2+x^4$ and an anharmonic oscillator $V(x)=x^2+x^4$. As it
can be seen from the figure, both plots start from zero and show a
semi-flat region for $L>L_c$ which contains the sought after
inflection point. At this point, the curvature of the curve
vanishes. A closer look at the curve shows that this point is close
to $L=3.5$ for $N=8$, $k=1$ and $\lambda=1$. Note that, there also
exists a shallow minimum which is due to $L$ being just large enough
to detect the minima. Since $L$ is too small in this case to see the
overall structure of the potential, this minimum should be ignored.
Moreover, the ground state energy of this model is $E_0\simeq0.66$
\cite{taseli2} which is approximately equal to the energy of the
inflection point in the left part of Fig.~\ref{NL}. Therefore, the
negative shallow minimum in the figure is not related to the actual
value of the energy level which is certainly positive.

To obtain more accurate results, we need to use more basis
functions. Table \ref{Tab1} shows the first six lowest energy levels
of a double-well potential ($k=1$) for a small $\lambda$ regime
($\lambda=0.01$). We have used $N=100$ basis functions in order to
obtain the near-exact energy eigenvalues. For this case, the ground
state energy is obtained with $82$ significant digits which we have
only reported it with $30$ significant digits. As it is expected,
using a fixed $N$, the ground state energy has the maximum precision
and the accuracy of the higher energy levels decrease in comparison
to the lower ones. To be more precise, the sixth eigenvalue has $79$
significant digits which still is very high. Moreover, the optimum
length scale $\hat L$ for $100$ basis functions is equal to
$16.70762$. This value strictly depends on the number of the basis
functions and will increase for a larger set of them. Since the
potential shifted by $1/(4\lambda)$ is positive definite, some
authors have reported the highly accurate positive definite
eigenvalues $E_n+1/(4\lambda)$ \cite{Banerjee}. As it can be seen
from Table \ref{Tab1}, our results are in complete agreement with
them. It is worth to mention that, using the periodic boundary
condition, the precision of 30 significant digits can be obtained
using $45$ basis functions. While, for Dirichlet boundary condition,
we need to use $55$ basis function to get the same accuracy
\cite{taseli2}. In fact, the losing of accuracy is due to the
mandatory vanishing of the wave function at $x=\pm L$. Note that, we
have also used the optimized value of $L$ ($\hat L$) for the ground
state energy to compute the higher energy levels $E_n$. In fact, we
have observed that $\hat L$ depends only on the number of basis
functions $N$ not on the south after energy levels. This is also
true for the case of Dirichlet boundary condition
\cite{taseli,taseli2,Bhattacharyya,Bhattacharyya2}. In principle, we
need to use more basis functions to obtain the same accuracy for the
higher energy levels. So, reduction of the accuracy of higher energy
eigenvalues (using fixed $N$) is natural and is not due to the
optimization procedure (Tables \ref{Tab1}-\ref{Tab3}). Although odd
basis functions $\phi_m^1(x)$ have the same form for both Dirichlet
and periodic boundary conditions, their optimized value of $L$ is a
little different which results in more accurate eigenvalues even for
odd states using periodic boundary condition.

\begin{figure}
\centering
  \includegraphics[width=8cm]{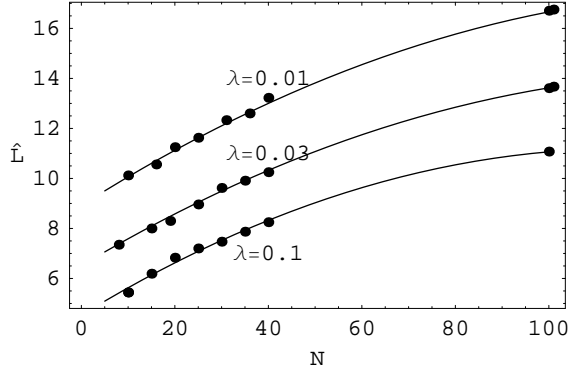}\\
  \caption{$\hat{L}$ versus $N$ for $k=1$, and
   $\lambda=\{0.01,0.03,0.1\}$. Circles show the position of the calculated inflection points and
   the curve is its interpolating function.}\label{NL2}
\end{figure}

For the larger values of $\lambda$, the accuracy of the results
increase even using the same number of basis functions (Table
\ref{Tab2} and \ref{Tab3}), whereas the optimum value of $L$
decreases. This is due to the decreasing of the classical turning
points for larger $\lambda$. These results are also remarkably
accurate in comparison with the near-exact values of
Refs.~\cite{Banerjee,Bhattacharyya3}. Figure \ref{NL2} shows the optimum value of
$L$ versus $N$ for $\lambda=0.01,0.03,0.1$. We can also use the
scaling properties of the double-well potential to find the
properties of a general Hamiltonian $H(k,\lambda)=-p^2-kx^2+\lambda
x^4$ from a more simpler one which we have studied here
$H(1,\beta)=-p^2-x^2+\beta x^4$. Using the transformation of
variable from $x$ to  $k^{1/4}x$, we find the following scaling
properties \cite{Banerjee,taseli2}
\begin{eqnarray}
H(k,\lambda)&=&k^{1/2} H(1,\beta), \\
E(k,\lambda)&=& k^{1/2}E(1,\beta), \\
\hat L(k,\lambda)&=&k^{-1/4}\hat L(1,\beta),
\end{eqnarray}
where $\beta=k^{-3/2}\lambda$. Thus, it is sufficient to consider
only the reduced potential ($k=1$).


\begin{table}
 \centering
  \caption{First six lowest energy eigenvalues of a double-well potential ($k=1$) for $\lambda=0.01$, $N=100$ and $\hat L=16.70762$. \emph{SD} denotes the
  number of significant digits.}\label{Tab1}
\begin{tabular}{ccc}\\\hline\hline
$n$&$E_n$&\footnotesize{\it{SD}}\\\hline
$0$&-23.5959513947022931175742924292&82 \\
$1$&-23.5959513947022931173974337194&81\\
$2$&-20.8298063940006898721661249287&80\\
$3$&-20.8298063940006897803867088013&80\\
$4$&-18.1299111662859753878276848315&78\\
$5$&-18.1299111662859531975740043181&78\\\hline\hline
\end{tabular}
\end{table}

\begin{table}
 \centering
\caption{First six lowest energy eigenvalues of a double-well
potential ($k=1$) for $\lambda=0.03$, $N=100$ and $\hat
L=13.60979$.}\label{Tab2}
\begin{tabular}{ccc}\\\hline\hline
$n$&$E_n$&\footnotesize{\it{SD}}\\\hline
$0$&-6.95073188927955191828148104931&97\\
$1$&-6.95072754950196756189760500468&95\\
$2$&-4.32728413386759375726086836212&94\\
$3$&-4.32667786658379381203893295176&94\\
$4$&-1.98615994840071249926930230256&92\\
$5$&-1.95646376927817057309963393657&92\\\hline\hline
\end{tabular}
\end{table}

\begin{table}
 \centering
  \caption{First six lowest energy eigenvalues of a double-well potential ($k=1$) for $\lambda=0.1$, $N=100$ and $\hat L=11.07433$.}\label{Tab3}
\begin{tabular}{ccc}\\\hline\hline
$n$&$E_n$&\footnotesize{\it{SD}}\\\hline
$0$&-1.26549283721398510854595401983&104 \\
$1$&-1.15305913107745006809098709688&104\\
$2$&0.509488545436203212948452569004&103\\
$3$&\,\,1.54354603976759862420138901373&103\\
$4$&\,\,3.10513379668314777728015050384&101\\
$5$&\,\,4.83611381900421025918208666909&101\\\hline\hline
\end{tabular}
\end{table}

\section{Conclusions}\label{sec4}
We have proposed the usage of periodic boundary condition on the
basis function for accurate calculation of eigenvalues and
eigenfunctions of a symmetric double-well potential. In this case,
the Rayleigh-Ritz variational formalism is not applicable. Because,
the graph of the energy versus the domain of the variable starts
from zero and grows as this domain increases. We showed that this
graph, because of the form of the potential, may contain some minima
which are not related to the actual value of the energy and should
be ignored. On the other hand, we exhibit that the behavior of the
energy versus the variable's domain shows the existence of an
inflection point with vanishing curvature which is related to the
optimum value of this domain. So, the variational procedure is
finding this inflection point by varying the domain of the variable.
We also showed that, for a fixed number of basis functions, the
results obtained from Dirichlet boundary condition
(particle-in-a-box) are less accurate in comparison to the usage of
periodic boundary condition. In fact, by using the particle-in-a-box
bases, we enforce the wave functions to vanish at the boundaries
which results in losing the accuracy. But, the periodic boundary
condition lets the wave functions to properly fit themselves to the
exact solutions. Therefore, we can get the same accuracy with a
smaller number of basis. For instance, for the case of $k=1$,
$\lambda=0.01$ and $N=45$ basis functions, we have obtained the
energy eigenvalues with $30$ significant digits. This accuracy, for
Dirichlet boundary condition, is obtainable using $N=55$
particle-in-a-box basis. In fact, this simple and efficient method
is rarely used by physicists to tackle eigenvalue problems and even
they sometimes use some complicated and inaccurate techniques
\cite{Monerat}. Although, we have studied only double-well
oscillators, but the presented method is applicable quite generally
for eigenvalue problems having polynomial potential which may
correspond to physically relevant situations \cite{pedram}.
Moreover, this method can be extended and used for two- or
three-dimensional cases. We expect that the priority of the periodic
boundary condition to the Dirichlet boundary condition remains also
in higher dimensions. In our future work, we will investigate this
issue and compare our results with the exact or near-exact ones
\cite{taseli3}.


\end{document}